\documentclass[10pt,aps,prd,twocolumn,showpacs,amsmath,amssymb,nofootinbib,eqsecnum,preprintnumbers,superscriptaddress]{revtex4-2}
\usepackage{amsmath,amssymb}
\usepackage{booktabs}   
\usepackage{multirow}   
\usepackage{array}      

\usepackage{amsmath,amssymb}
\usepackage{enumitem}  
\usepackage[usenames, dvipsnames]{color} 
\usepackage{graphicx} 
\usepackage{comment}
\usepackage[normalem]{ulem}
\usepackage[utf8]{inputenc}
\usepackage{url}
 \usepackage{xcolor}
 \usepackage{physics}

\usepackage{hyperref}  

\usepackage{graphicx}
\usepackage{dcolumn}
\usepackage{bm}

\newcommand{\be}{\begin{equation}}
\newcommand{\ee}{\end{equation}}
\newcommand{\ba}{\begin{eqnarray}}
\newcommand{\ea}{\end{eqnarray}}

\newcommand{\beq}{\begin{equation}}
\newcommand{\eeq}{\end{equation}}
\newcommand{\beqa}{\begin{eqnarray}}
\newcommand{\eeqa}{\end{eqnarray}}

\newcommand{\eps}{\varepsilon}

\newcommand{\A}[1]{A^{\!(#1)}}                 

\newcommand{\dg}{{{n}}}                           

\begin{document}

\title{On Generalized (Conformal) Killing Tensors}

\author{Cynthia Arias}
\email{cynthia.arias@matfyz.cuni.cz}

\affiliation{Institute of Theoretical Physics, Faculty of Mathematics and Physics,
Charles University, Prague, V Hole{\v s}ovi{\v c}k{\' a}ch 2, 180 00 Prague 8, Czech Republic}

\author{David Kop{\v c}an}

\email{david.kopcan564@student.cuni.cz}

\affiliation{Institute of Theoretical Physics, Faculty of Mathematics and Physics,
Charles University, Prague, V Hole{\v s}ovi{\v c}k{\' a}ch 2, 180 00 Prague 8, Czech Republic}

\author{David Kubiz\v n\'ak}

\email{david.kubiznak@matfyz.cuni.cz}

\affiliation{Institute of Theoretical Physics, Faculty of Mathematics and Physics,
Charles University, Prague, V Hole{\v s}ovi{\v c}k{\' a}ch 2, 180 00 Prague 8, Czech Republic}

\author{Marek Mili{\v c}ka}
\email{marek.milicka@gmail.com }


\affiliation{Institute of Theoretical Physics, Faculty of Mathematics and Physics,
Charles University, Prague, V Hole{\v s}ovi{\v c}k{\' a}ch 2, 180 00 Prague 8, Czech Republic}

\date{July 21, 2026}

\begin{abstract}
We study higher-rank generalized Killing tensors with mixed symmetries, providing a couple of examples and applications. It is shown that such objects naturally exist in higher-dimensional rotating black hole spacetimes, where they arise as partially contracted `squares' of Killing--Yano tensors and display interesting algebraic and differential properties. Motivated by conformal Killing tensors, a generalization of these objects that gives rise to parallel-transported tensors along null geodesics is proposed and shown to exist in higher-dimensional rotating black hole spacetimes. 
\end{abstract}

 \maketitle

\section{Introduction}

Symmetries are the foundations of modern physics. They simplify complex equations, prescribe the fundamental forces of nature, and directly lead to the conservation laws that govern physical reality. In general relativity, the simplest (continuous global) symmetries correspond to the isometries of the metric. They are described by Killing vector fields $\xi^a$, obeying the {\em Killing vector} field equation:
\be 
\nabla_{(a}\xi_{b)}=0\,.
\ee 
Such symmetries, however, are just the tip of the iceberg of all possible symmetries. In this paper, we focus on the so-called {\em hidden symmetries} that directly generalize Killing vectors and are encoded in higher-rank tensor fields.

Among such hidden symmetries, perhaps the best known and well-understood are 
the completely 
symmetric {\em Killing tensors} \cite{staeckel1895integration, Walker:1970un}: 
\be \label{KT}
K_{a_1\dots a_p}=K_{(a_1\dots a_p)}\,,\quad 
\nabla_{(a}K_{a_1a_2\dots a_p)}=0\,,
\ee 
and the completely antisymmetric 
{\em Killing--Yano (KY) forms} \cite{yano1952some, Penrose:1973um}:
\be\label{KY} 
f_{a_1\dots a_p}=f_{[a_1\dots a_p]}\,,\quad 
\nabla_{(a}f_{a_1)a_2\dots a_p}=0\,.
\ee 
Both are extremely useful, especially in rotating black hole spacetimes, where 
they complement Killing vectors, giving rise to integrable {\em geodesic motion} and generating 
{\em symmetry operators} for various test field equations that allow for their {\em separability}, e.g., \cite{Carter:1968rr, Carter:1977pq, Carter:1979fe, Frolov:2017kze}.

However, as noted by Collinson and Howarth \cite{collinson2000generalized}, when it comes to {\em parallel transport}, there is no a priori reason to restrict to completely symmetric or antisymmetric objects, and one can, in principle, employ higher-rank tensors with mixed symmetries, giving rise to a notion of {\em generalized Killing tensors (GKTs).}

It is the aim of this paper to pick up the threads of these developments. First, we further relax the restrictions on admissible symmetries of GKTs, imposing no symmetry whatsoever on such objects. This naturally leads to 
{\em equivalence  classes} of symmetry objects,  
whose {\em representatives} 
coincide upon the due symmetrization of indices and all lead to the same parallel-transported objects.  
We then focus on some special cases of such tensors and establish several of their properties. We also show that such objects naturally exist in higher-dimensional rotating black hole spacetimes, encoded in the {\em Kerr-NUT-AdS geometries} \cite{Chen:2006xh, Frolov:2017kze}, thus providing the (first recorded) non-trivial example of the developed theory. We also point out that they have applications in several physical situations, including the recently studied higher-rank generalized {\em Proca theories} \cite{Kubiznak:2026vlq, Kubiznak:2026snm}.

Our paper is organized as follows. In 
Sec.~\ref{Sec2}, we review the GKTs of Collinson and Howarth and introduce their non-symmetric versions. In Sec.~\ref{Sec3}, we focus on a special case of such tensors and study their properties. Non-trivial examples of these objects in higher-dimensional rotating black hole spacetimes are presented in Sec.~\ref{Sec4}. In Sec.~\ref{Sec5}, motivated by conformal Killing tensors, we attempt to extend GKTs to the conformal case, showing that while a useful weaker version of GKTs can be defined, the corresponding conformal property seems to be lost. We conclude in Sec.~\ref{Sec6} with a summary and numerous questions for future studies. Appendix~\ref{AppA}   
is devoted to the review of (conformal) KY tensors and their parallel transport properties. Appendix~\ref{AppB} contains a technical calculation of the construction of conformal GKTs from a square of closed conformal KY $p$-forms used in the main text.

\section{Generalized Killing tensors}
\label{Sec2}

In this section, we will review
the GKTs of Collinson and Howarth \cite{collinson2000generalized} and introduce their non-symmetric versions.

\subsection{GKTs of Collinson and Howarth}

Motivated by parallel transport, Collinson and Howarth  \cite{collinson2000generalized} (see also \cite{nikitin2005generalized}) defined the following objects characterized by two types of indices: $q$ completely symmetric indices $\{s_1,\dots, s_q\}$ and $p$ additional indices $\{a_1,\dots, a_p\}$ with no symmetries imposed, forming together a  `{\em rank-($p$-$q$)}' $B$-tensor:
\be\label{B_CH} 
B_{a_1\dots a_p s_1\dots s_q}=B_{a_1\dots a_p (s_1\dots s_q)}\,,
\ee 
obeying the following equation:
\be\label{Beq} 
\nabla_{(a}B_{|a_1\dots a_p| s_1\dots s_q)}=0\,.
\ee 
We shall call the objects \eqref{Beq} obeying, in addition, the symmetry requirement \eqref{B_CH} the {\em GKTs of Collinson and Howarth} \cite{collinson2000generalized}  of rank-($p$-$q$).

For such tensors, there is a 1-1 correspondence with the following 
objects: 
\be\label{wB} 
w_{a_1\dots a_p}\equiv B_{a_1\dots a_p s_1\dots s_q} u^{s_1}\dots u^{s_q}\,,
\ee 
that are parallel-transported along any (affine-parametrized) geodesic 
\be \label{PTproof}
\nabla_u u^a=u^b \nabla_b u^a=0\,.
\ee 
Indeed, using \eqref{wB} and \eqref{Beq}, we have
\ba 
\nabla_u w_{a_1\dots a_p}&=&u^a\nabla_a(B_{a_1\dots a_p s_1\dots s_q} u^{s_1}\dots u^{s_q})\nonumber\\
&=&u^au^{s_1}\dots u^{s_q}\nabla_a B_{a_1\dots a_p s_1\dots s_q}\nonumber\\
&=&
u^au^{s_1}\dots u^{s_q}\nabla_{(a} B_{|a_1\dots a_p| s_1\dots s_q)}=0\,.\qquad \label{PTGKTs}
\ea

Note that Killing tensors are recovered as objects where only symmetric indices are present (that is, all $a$ indices vanish), and KY tensors as objects that are completely antisymmetric and have one $s$ index. 
Note also that by contracting the $a$ indices with the (covariantly constant) metric, we can construct lower-rank GKTs of Collinson and Howarth, e.g. 
\be 
B_{a_3\dots a_ps_1\dots s_q}\equiv B^{c}{}_{c a_3\dots a_{p}s_1\dots s_q}\,.
\ee 
In particular, when all $a$-indices get contracted, we recover the standard Killing tensors obeying \eqref{KT}. 

\subsection{GKTs with no extra symmetries}
While Collinson and Howarth required $B$ to be fully symmetric in the $s$-indices, this condition is stronger than what parallel transport actually requires. We therefore consider a more general object:
\be 
B_{a_1\dots a_p b_1\dots b_q}\,,
\ee 
where no symmetry is imposed on either the $b$-indices or the $a$-indices, requiring only that it satisfies the {\em GKT equation}:
\be\label{GKT} 
\nabla_{(a}B_{|a_1\dots a_p| b_1\dots b_q)}=0\,. 
\ee 
We call such objects  {\em GKTs of rank-($p$-$q$)}, to distinguish them from the Collinson and Howarth case; we shall also sometimes refer to the $b$ indices as the `parallel-transported indices'.

It follows that the objects 
\be\label{wB2} 
w_{a_1\dots a_p}\equiv B_{a_1\dots a_p b_1\dots b_q} u^{b_1}\dots u^{b_q}\,,
\ee 
are still parallel-transported along any affine parametrized geodesic $u^a$. This is because the contraction with the symmetric product $u^{b_{1}}...u^{b_{q}}$ effectively symmetrizes $B$ over its $b$-indices, and so Eq.~(\ref{PTGKTs}) remains unchanged. Note, however, that the map between the new $B$-objects and the corresponding parallel-transported tensors $w$, \eqref{wB2}, is no longer 1-1; only a symmetric part of the $b$-indices contributes to $w$.

This observation motivates the definition of equivalence classes of GKTs. {Namely, consider} two tensors $B$ and $B'$ of the same rank-($p$-$q$), satisfying \eqref{GKT}. Then, they belong to the same equivalence class, $B\sim B'$, if and only if they give rise to the same parallel-transported objects, $w=w'$, if and only if they coincide upon complete symmetrization of their 
$b$-indices:
\be
B\sim B'\quad \Leftrightarrow \quad B_{a_{1}...a_{p}(b_{1}...b_{q})}=B^{\prime}_{a_{1}...a_{p}(b_{1}...b_{q})}\,.
\ee
Each equivalence class contains a unique representative with the $b$-indices fully symmetrized, which is precisely the GKT of Collinson and Howarth. The correspondence with the parallel-transported objects is 1-1 
only at this symmetric level. Once the symmetry is relaxed, $B$ $\mapsto$ $w$ is a `many-to-one' map, yielding the same parallel-transported $w$. This is exactly the freedom that the Collinson--Howarth construction eliminates, {but we} wish to retain.  This will allow us to consider representatives with various additional symmetries that 
are physically or geometrically `preferred'.\footnote{Of course,
any tensor that is totally antisymmetric in the $b$-indices with  $q\geq 2$ automatically satisfies the GKTs equation \eqref{GKT}. Since it has a vanishing symmetric $b$-part, it would not serve as a representative of a non-trivial parallel-transported object.
}
In particular, 
as we shall see in Sec.~\ref{Sec3}, the construction of GKTs from products of KY tensors `naturally' leads to `Riemann-type' representatives with Riemann-type symmetries. While passing to the Collinson--Howarth form is always possible by symmetrization, it generically destroys additional symmetries carried by the non-symmetrized objects.

Note also that 
while any representative of the equivalence class suffices for parallel transport, specific representatives may be preferred for other applications, such as the construction of symmetry operators for the test field equations. 
Whether this is indeed the case goes beyond the scope of the current paper and will be studied elsewhere.

`Redundant' GKTs can be obtained by taking direct products of lower-rank GKTs. Namely, having a rank-$(p_A$-$q_A)$ GKT $A$ and 
a rank-$(p_B$-$q_B)$ GKT $B$, their product
\ba 
C_{a_1\dots a_{(p_A+p_B)}b_{1}\dots b_{(q_A+q_B)}} \equiv A_{a_1\dots a_{p_A}b_1\dots b_{q_A}}\qquad\quad\nonumber\\
\times B_{a_{(p_A+1)}\dots  a_{(p_A+p_B)}b_{(q_A+1)}\dots b_{(q_A+q_B)}}
\quad 
\ea 
is again a rank-$((p_A+p_B)-(q_A+q_B))$ GKT. Following the literature on Killing tensors, e.g. \cite{Walker:1970un}, we call GKTs that can be obtained as linear combinations of such products {\em reducible} GKTs. On the other hand, when it is not possible to decompose a GKT in such a way, we call it {\em irreducible}.

\section{Riemann-type generalized Killing tensors}\label{Sec3}

Let us now focus on a particular case of the rank-(2-2) GKTs. When such objects arise as squares of KY tensors, 
they inherit Riemann-type symmetries and go beyond the original GKT definition of Collinson and Howarth. As we will see in the next section, such `Riemann-type GKTs' naturally exist in higher-dimensional rotating black hole spacetimes.

\subsection{Rank-(2-2) GKTs}
Let us focus on the following rank-(2-2) tensor $Q_{abcd}$, obeying the GKT equation
\be\label{R1} 
\nabla_{(a}Q_{b|c|d)e}=0\,,
\ee 
with no additional symmetries imposed on $Q$. This is a special case of the (not necessarily symmetric) GKT, where we `placed' the parallel-transported indices in the first and third slots. 
Of course, imposing the additional symmetry, namely 
\be\label{R12} 
Q_{abcd}=Q_{(a|b|c)d}\,,
\ee 
would make $Q$ into a GKT of Collinson and Howarth; we shall not impose this symmetry at the moment. 

Obviously, 
\be 
Q_{ac}\equiv g^{bd}Q_{abcd}=Q_{abc}{}^b
\ee 
is then a standard (not necessarily symmetric) Killing tensor, obeying 
\be \label{eqQab}
\nabla_{(a}Q_{bc)}=0\,.
\ee

According to \eqref{wB}, object $Q_{abcd}$ automatically generates a rank-2  tensor
\be \label{Qu}
w_{bd}\equiv Q_{abcd}u^{a}u^c\
\ee 
that is parallel-transported 
along geodesics $u^a$. Conversely, parallel-transported objects of the form \eqref{Qu} give rise to an equivalence class of rank-(2-2) GKTs $Q_{abcd}$.

\subsection{Riemann-type GKTs}

A particular example of the GKT tensor can be obtained by a contracted product of two KY $p$-forms, $f^{(1)}$, and $f^{(2)}$, namely: 
\be\label{RiemanntypeGKTS}
Q_{abcd}\equiv \frac{1}{2}\left(f^{(1)}_{ab a_3\dots a_p}f^{(2)}_{cd}{}^{a_3\dots a_p}+f^{(1)}_{cd}{}^{a_3\dots a_p}f^{(2)}_{ab a_3\dots a_p}\right)\,.
\ee 
That this is a GKT can be seen as follows. Consider the following $(p-1)$-forms (c.f. \eqref{w_KY}):
\be 
w^{(1)}_{a_2\dots a_p}\equiv u^a f^{(1)}_{aa_2\dots a_p}\,,\quad w^{(2)}_{a_2\dots a_p}\equiv u^a f^{(2)}_{aa_2\dots a_p}\,.
\ee 
Such forms are parallel-transported; namely, we have 
\be 
\nabla_u w^{(1)}_{a_2\dots a_p}=u^b u^a\nabla_b f^{(1)}_{aa_2\dots a_p}=
u^b u^a\nabla_{(b} f^{(1)}_{a)a_2\dots a_p}=0\,.
\ee 
on behalf of Eq.~\eqref{KY}, and similarly for $w^{(2)}$.
This means that any product of $w^{(1)}$ and $w^{(2)}$, and their contractions, will be parallel transported. In particular, consider
\ba\label{wab}
w_{ab}&\equiv&\frac{1}{2}\Bigl(
w^{(1)}_{aa_3\dots a_p} w^{(2)}{}_b{}^{a_3\dots a_p}+w^{(1)}_{ba_3\dots a_p} w^{(2)}{}_a{}^{a_3\dots a_p}\Bigr)\nonumber\\
&=&Q_{cadb}u^{c}u^d\,,
\ea 
where one representative of $Q_{abcd}$ is given by \eqref{RiemanntypeGKTS}.\footnote{Note that while $w_{ab}$ in \eqref{wab} is symmetric in $a$ and $b$ indices, the corresponding $Q_{cadb}$ need not be. In fact, demanding such symmetry would yield trivial $Q_{cadb}$.} Since this is of the form \eqref{Qu}, $Q_{abcd}$ constructed in \eqref{RiemanntypeGKTS} must be a GKT, obeying \eqref{R1}. 

By construction \eqref{RiemanntypeGKTS},
the above GKT has the following additional symmetries:
\be \label{RT1}
Q_{abcd}=Q_{[ab]cd}=Q_{ab[cd]}\,,
\ee 
and  
\be \label{RT2}
Q_{abcd}=Q_{cdab}\,.
\ee 
We call a GKT tensor with both such symmetries a {\em Riemann-type GKT.} Note, however, that this tensor need not satisfy the Bianchi identities for the Riemann tensor, $R_{a[bcd]}=0$ and $R_{ab[cd;e]}=0$.

Note also that such a tensor cannot be a GKT of Collinson and Howarth, since demanding \eqref{R12} in addition to the Riemann-type symmetries \eqref{RT1} would result in a vanishing tensor: indeed, we would have  
\ba
Q_{abcd} &=& Q_{cbad} =- Q_{cbda} = -Q_{dbca}\nonumber\\
&=&Q_{dbac}=Q_{abdc}=-Q_{abcd}=0\,.   
\ea
At the same time, defining 
\be 
\tilde Q_{abcd}\equiv Q_{(a|b|c)d}\,,
\ee 
would yield a GKT of Collinson and Howarth. The price to pay is that the `lucrative' Riemann-like symmetries \eqref{RT1} would no longer be satisfied, and we would only preserve  \eqref{RT2}. 

Let us also note that the Riemann-type GKTs have recently found an application in generalized higher-rank form {\em Proca-type theories} \cite{Colladay:1998fq, Heisenberg:2014rta, Feng:2015sbw, Fan:2017bka, Kubiznak:2026snm}, where they naturally appear in stealth solutions generated from 
KY $p$-forms \cite{Kubiznak:2026snm}.

\subsection{Reducible Riemann-type GKT}

When the Riemann-type Killing tensor $Q_{abcd}$, given by \eqref{RiemanntypeGKTS} specialized to $p=2$, is obtained by a product of two copies of a rank-2 KY tensor $f_{ab}$ (no contractions take place):
\be \label{Qff2}
Q_{abcd}=f_{ab}f_{cd}\,, 
\ee 
it is reducible and {\em trivial} in the following sense. Let us construct the corresponding $w_{ab}$,
\be 
w_{ab}=f_{ac}f_{bd}u^c u^d=w_aw_b\,, \quad w_a=f_{ac}u^c\,.
\ee
Since $w_{ab}$ is a  product of the same vector with itself, it has exactly one non-trivial eigenvalue, $w_aw^a$. This is the same constant of motion $Q_{ab}u^au^b$ as obtained from the corresponding trace Killing tensor 
$Q_{ab}=Q_{acb}{}^c$.   

On the other hand, considering a higher-rank KY $p$-form, its partially contracted product $Q_{abcd}$ is no longer trivial; it is irreducible, and the  resultant $w_{ab}$ has several non-trivial eigenvalues.

\subsection{Differential identities}

Requiring certain  symmetries of the GKTs yields interesting differential identities. Namely, if $Q_{abcd}$ satisfies the Collinson and Howarth symmetry \eqref{R12}, or has the Riemann-type symmetry \eqref{RT2}, then the associated $Q_{ab}=Q_{acb}{}^c$ is symmetric  and \eqref{eqQab} reads
\be\label{eqQsym}
\nabla_a Q_{bc} + \nabla_b Q_{ac} + \nabla_c Q_{ab} = 0\,.
\ee 
Taking the trace, we get
\be\label{Phi1} 
\nabla_a Q^{ab}=-\tfrac{1}{2}\nabla^b Q^a{}_a\,.
\ee

Similarly, taking the $ab$ trace of 
\be\label{Phi2} 
\nabla^{(a}Q^{b|c|d)e}=0\,,
\ee
we get
\begin{align}
&\nabla^{a}Q_{a}{}^{cde} + \nabla^{a}Q^{dc}{}_{a}{}^{e} + \nabla^{b}Q_{b}{}^{cde}\nonumber\\
&+ \nabla^{b}Q^{dc}{}_{b}{}^{e} + \nabla^{d}Q_{a}{}^{cae} + \nabla^{d}Q^{ac}{}_{a}{}^{e} = 
0\,.
\end{align}
In the Collinson--Howarth case, the last two terms do not admit an obvious geometric interpretation; by contrast, under the Riemann-type symmetries they reduce to gradients of $Q_{ab}$. 
Using these symmetries, the equation can be rewritten as
\be
\nabla_a Q^{a(b|c|d)}=-\tfrac{1}{2}\nabla^c Q^{bd}.
\ee
In the same fashion, taking the $ac$ trace and using the Riemann symmetries, we get
\be\label{Phi3} 
\nabla_a Q^{(b|a|d)e}=\nabla^{(b}Q^{d)e}=-\tfrac{1}{2}\nabla^e Q^{bd}\,,
\ee 
where the last equality follows from first using the relation \eqref{eqQsym}.
These identities found their applications for constructing stealth solutions of generalized Proca theories in \cite{Kubiznak:2026snm}.

\subsection{Higher-rank generalizations}

A generalization of \eqref{RiemanntypeGKTS} to higher-rank tensors, such as 
\be 
Q_{abcdef}=
\frac{1}{2}(f^{(1)}_{abc a_4\dots a_p}f^{(2)}_{def}{}^{a_4\dots a_p}+f^{(1)}_{def}{}^{a_4\dots a_p}f^{(2)}_{abc a_4\dots a_p})\,,
\ee 
automatically obeying 
\be 
\nabla_{(a}Q_{b|a_1a_2|c)a_3a_4}=0\,,
\ee 
is obvious. It has the following symmetries:
\be 
Q_{abcdef}=Q_{[abc]def}=Q_{abc[def]}=Q_{defabc}\,,
\ee 
and similarly for higher-rank GKTs of this type. When non-trivial contractions take place, such GKTs are irreducible.

\section{Example}\label{Sec4}

When Collinson and Howarth defined their GKTs \cite{collinson2000generalized}, they provided only (not very interesting) examples in spherically symmetric spacetimes. It is the aim of this section to show that GKTs naturally appear in higher-dimensional rotating black hole spacetimes, encoded in the {\em Kerr-NUT-AdS geometry} \cite{Chen:2006xh, Frolov:2017kze}.

\subsection{Kerr-NUT-AdS spacetime}

Parameterizing the spacetime dimension ${D=2\dg+\eps}$ (with $\eps=0$ in even and $\eps=1$ in odd dimensions), the general Kerr-NUT-AdS geometry can be written as follows  \cite{Chen:2006xh, Frolov:2017kze}:
\ba\label{KerrNUTAdSmetric}
ds^2&=&\sum_{\mu=1}^\dg\;\biggl[\; \frac{U_\mu}{X_\mu}\,{dx_{\mu}^{2}}
  +\, \frac{X_\mu}{U_\mu}\,\Bigl(\,\sum_{j=0}^{\dg-1} \A{j}_{\mu}d\psi_j \Bigr)^{\!2}
  \;\biggr]\nonumber\\
  &&  +\eps\frac{c}{\A{\dg}}\Bigl(\sum_{k=0}^\dg \A{k}d\psi_k\!\Bigr)^{\!2}\;,
\ea
where 
the functions ${\A{k}}$, ${\A{j}_\mu}$, and ${U_\mu}$ are `symmetric polynomials' of coordinates ${x_\mu}$:
\ba\label{AUdefs} \A{k}&=&\!\!\!\!\!\sum_{\substack{\nu_1,\dots,\nu_k=1\\\nu_1<\dots<\nu_k}}^\dg\!\!\!\!\!x^2_{\nu_1}\dots x^2_{\nu_k}\;,
\qquad
  \A{j}_{\mu}=\!\!\!\!\!\sum_{\substack{\nu_1,\dots,\nu_j=1\\\nu_1<\dots<\nu_j\\\nu_i\ne\mu}}^\dg\!\!\!\!\!x^2_{\nu_1}\dots x^2_{\nu_j}\;,\nonumber\\
U_{\mu}&=&\prod_{\substack{\nu=1\\\nu\ne\mu}}^\dg(x_{\nu}^2-x_{\mu}^2)\;,
\ea
$c$ is a constant that appears in odd dimensions, 
and each metric function ${X_\mu}$ is a function of a single coordinate ${x_\mu}$:
\begin{equation}\label{Xfcdependence}
    X_\mu=X_\mu(x_\mu)\;.
\end{equation}
When the vacuum Einstein equations with a cosmological constant are imposed, 
$X_\mu$ take the following explicit form:
\be 
X_\mu=
\begin{cases}
-2b_\mu x_\mu+\sum_{k=0}^n c_k (x_\mu)^{2k}\quad D\ \mbox{is even}\\
-\frac{c}{x_\mu^2}-2b_\mu+\sum_{k=1}^n c_k (x_\mu)^{2k}\quad D\ \mbox{is odd}\,.
\end{cases}
\ee 
This, in particular, describes higher-dimensional multiply-spinning Myers--Perry black holes \cite{Myers:1986un} or the Kerr-AdS spacetimes in all dimensions \cite{Gibbons:2004js, Gibbons:2004uw}; see \cite{Frolov:2017kze} for specific details and various subcases.

Interestingly, the described (hidden) symmetries (see below) are present irrespective of the field equations, and exist for a general class of the {\em off-shell} spacetimes characterized by \eqref{Xfcdependence}.

\subsection{Killing--Yano symmetries}

The general off-shell Kerr-NUT-AdS spacetime admits an  
extended tower of conformal KY symmetries (see Appendix~\ref{AppA} for a review of various types of KY tensors and their properties). 
Among these, the most important is the 
{\em principal tensor} $h$. This is a non-degenerate closed conformal KY 2-form $h_{ab}$, obeying 
\be 
h_{ab}=h_{[ab]}\,,\quad 
\nabla_a h_{bc}=g_{a[b}\xi_{c]}\,.
\ee 
From this definition, it follows that 
\be 
dh=0\,,\quad \xi_b=\frac{2}{D-1}\nabla^a h_{ab}\,.
\ee 
Moreover, being closed, $h$ can locally be expressed  as the exterior derivative of a one-form potential $b$:
\be 
h=db\,,\quad 
b=\frac{1}{2}\sum_{k=0}^{n-1} A^{(k+1)} d\psi_k\,.\label{PKYb}
\ee 
Since a wedge product of two closed conformal KY tensors is again a closed conformal KY tensor \cite{Krtous:2006qy}, the principal tensor 
generates a tower of closed conformal KY $(2p)$-forms, given by 
\be \label{hp}
h_1\equiv h\,,\quad h_p\equiv \underbrace{h\wedge \dots\wedge h}_{\mbox{\small $p$-times}}\,. 
\ee 
Their Hodge duals are the rank-$(D-2p)$ KY tensors
\be 
f_{p}=*h_p\,.
\ee 
In even (odd) dimensions, these have even (odd) rank.

\subsection{Killing tensors and GKTs}

When $(f_p)$'s are `squared', so that two indices are left non-contracted, they give rise to the following Killing tensors \cite{Frolov:2017kze}: 
\be \label{Qpab}
Q_{ab}^{(p)}\equiv(f_p)_{ab_2\dots b_{D-2p}}(f_p)_b{}^{b_2\dots b_{D-2p}}\,.
\ee 
In $D=2n+\eps$, $\eps = 0,1$, spacetime dimensions, there are ($n-1$) irreducible tensors of this type, giving rise to $(n-1)$ Carter-like constants of geodesic motion. Together with the additional $(D-n+1)$ constants coming from the metric and its isometries  $\partial_{\psi_j}$ ($j=0,\dots n-1+\eps$), this yields completely integrable geodesic motion in these spacetimes 
\cite{Page:2006ka, Krtous:2006qy}. Such tensors also allow for the separability of the Klein--Gordon equation \cite{Frolov:2006pe, Sergyeyev:2007gf} and play a role in the separability of vector perturbations \cite{Krtous:2018bvk, Houri:2019lnu}.

When more indices are left non-contracted in the squares of the Killing--Yano tensors $f_p$, we recover the Riemann-type GKTs defined above:
\be\label{Qpabcd} 
Q_{abcd}^{(p)}\equiv(f_p)_{abb_3\dots b_{D-2p}}(f_p)_{cd}{}^{b_3\dots b_{D-2p}}\,.
\ee 
There are $(n-2+\eps)$ irreducible such objects, each of which gives rise to a (symmetric) rank-2 parallel-transported tensor $w_{ab}$, \eqref{wab}. It remains to be seen whether such objects play any direct role in the separability of (potentially higher-spin) test field equations. However, as described above, we already know that such objects play a natural role in the construction of $p$-form stealth solutions in higher-dimensional rotating black hole spacetimes  \cite{Kubiznak:2026snm}.

Similarly, one could also construct higher-rank Riemann-type GKTs. For example, we have
\be 
Q_{abcdef}^{(p)}\equiv (f_p)_{abcb_4\dots b_{D-2p}}(f_p)_{def}{}^{b_4\dots b_{D-2p}}\,.
\ee 
There are $(n-2)$ irreducible such objects in $D$ dimensions.

\subsection{Conformal Killing tensors}

Let us also note that apart from full Killing tensors constructed from $(f_p)$'s, \eqref{Qpab}, one also has a weaker structure of {\em conformal Killing tensors}, constructed from $(h_p)$'s, namely 
\be 
\tilde Q_{ab}^{(p)}\equiv (h_p)_{ab_2\dots b_{2p}}(h_p)_b{}^{b_2\dots b_{2p}}\,.
\ee 

A {\em conformal Killing tensor} of rank-2, $\tilde Q_{ab}$, obeys the following 
equation:\footnote{More generally, a conformal Killing tensor of rank $q$ is a tensor 
obeying 
\be 
\nabla_{(a}\tilde Q_{a_1\dots a_q)}=g_{(aa_1}\tilde Q_{a_2\dots a_q)}\,.
\ee 
As described in Sec.~\ref{Sec2}, in this definition of a conformal Killing tensor (and the same for \eqref{CKT}), we do not assume any additional complete symmetry (or tracelessness), as traditionally required in the literature, e.g., \cite{Walker:1970un}. 
Moreover, in this paper, we will entirely focus on rank-2 conformal Killing tensors and their GKT generalizations.
}
\be\label{CKT} 
\nabla_{(a}\tilde Q_{bc)}=g_{(ab}\tilde Q_{c)}\,,
\ee 
where $\tilde Q_c$ can be obtained by a contraction of the previous equation and reads:
\be 
\tilde Q_a= \frac{1}{D+2}\Bigl(\nabla_c\tilde Q^c{}_a+\nabla_c\tilde Q_a{}^c+\nabla_a \tilde Q_c{}^c\Bigr)\,.
\ee 
Such 
objects naturally give rise to constants for null geodesic motion:
\be\label{lgeo} 
\nabla_l l^a=0\,,\quad l^2=0\,,
\ee 
namely 
\be 
\tilde Q^{(l)}=\tilde Q_{ab}l^al^b\,. 
\ee 
Indeed, we have 
\ba 
\nabla_l \tilde Q^{(l)}&=&l^a l^b \nabla_l\tilde Q_{ab}=l^a l^b l^c \nabla_c\tilde Q_{ab}\nonumber\\
&=&l^a l^b l^c \nabla_{(c}\tilde Q_{ab)}=0\,,
\ea 
where, in the first step, we have used 
\eqref{lgeo}, and in the last step, we have employed the conformal Killing tensor equation \eqref{CKT} together with $l^2=0$.

The aim of the next section is to generalize the conformal Killing tensor equation \eqref{CKT} to GKTs. We focus on rank-(2-2) GKTs.

\section{On conformal generalization of  GKTs}\label{Sec5}

\subsection{Proposal}

Inspired by the conformal Killing tensor equation \eqref{CKT}, let us demand  
\be\label{CGKT} 
\nabla_{(a}  \tilde Q_{b|c|d)e}=g_{(ab}\xi_{|c|d)e}\,,
\ee 
and call an object obeying it a (2-2) {\em conformal-like generalized Killing tensor (CGKT)}.\footnote{As we shall see below, such objects do not really have a conformal symmetry under conformal transformations of the metric.} 
By contracting this equation, we recover 
\be 
\xi_{cde}=\frac{1}{D+2}(\nabla_d \tilde Q_{ce}+\nabla^b \tilde Q_{bcde}+\nabla^b \tilde Q_{dcbe})\,,
\ee 
where
\be 
\tilde Q_{ab}\equiv \tilde Q_{acb}{}^c\,.
\ee 
Obviously, on behalf of Eq.~\eqref{CGKT}, such a $\tilde Q_{ab}$ is necessarily a  conformal Killing tensor, obeying \eqref{CKT}, with $\tilde Q_c=\xi^b{}_{cb}$.

Moreover, considering the following 2-tensor:
\be 
F_{bd}\equiv Q_{abcd}l^al^c\,,
\ee 
it is automatically parallel-transported along null geodesics obeying \eqref{lgeo}. Indeed, we have 
\be 
\nabla_l F_{bd}=l^a l^c \nabla_l Q_{abcd}
=l^a l^c l^e \nabla_{(e} Q_{a|b|c)d}=0\,,
\ee  
on behalf of \eqref{CGKT}. 

\subsection{Finding the `right square root'}

Having proposed a conformal generalization of GKTs \eqref{CGKT}, one may wonder if it is possible to find examples of such tensors by squaring conformal KY tensors (see Appendix~\ref{AppA} for a review), generalizing the formula \eqref{RiemanntypeGKTS}.

That this may not be so simple can be seen as follows. 
Consider a product of two conformal KY 2-forms:
\be\label{kkill} 
Q_{abcd}=k_{ab}k_{cd}\,.
\ee 
Then,
\ba
\nabla_a Q_{bcde}&=&\nabla_a (k_{bc} k_{de})\nonumber\\
&=&(\nabla_a k_{bc}) k_{de} + k_{bc} (\nabla_a  k_{de})\nonumber\\ 
&=&(g_{a[b} \xi_{c]}+\nabla_{[a} k_{bc]} ) k_{de} +  k_{bc}(g_{a[d} \xi_{e]}+\nabla_{[a} k_{de]})\nonumber\\
&=& \frac{1}{2}(g_{ab} \xi_c -g_{ac} \xi_b) k_{de} + \frac{1}{2}(g_{ad} \xi_e -g_{ae} \xi_d) k_{bc} \nonumber\\
& \, & + \nabla_{[a} k_{bc]} k_{de} + k_{bc} \nabla_{[a} k_{de]}\,.
\ea 
Applying the symmetrization,
we find 
\ba
\nabla_{(a} Q_{b|c | d )e}
&=& \frac{1}{2}\Bigl( \xi_c g_{(ab} k_{d)e}  + \xi_e g_{(ad} k_{b)c}\nonumber\\
&&- g_{c(a} \xi_b  k_{d)e}
 -g_{e(a} \xi_d k_{b)c}
\Bigr)\,.
\ea
Obviously, whereas the first two terms have the right structure, the last two terms do not cancel and break the form of \eqref{CGKT};  thus \eqref{kkill} is not a CGKT in the sense of Eq.~\eqref{CGKT}.

To find the `right square root' of \eqref{CGKT}, let us proceed as follows. 
Consider first two closed conformal KY 2-forms $h$ and $k$.  They obey (see Eq.~\eqref{CCKYdef2})
\be 
\nabla_a h_{bc}=g_{a[b}\xi_{c]}\,,\quad {\nabla_a}
k_{bc}=g_{a[b}\eta_{c]}\,,
\ee 
and give rise to the two following parallel-transported 3-forms along any geodesic $u^a$ (see Eq.~\ref{FCCKY}):
\be 
F_{abc}=u_{[a}h_{bc]}\,,\quad 
H_{abc}=u_{[a}k_{bc]}\,.
\ee 
It means that any object constructed from $F, H$ and the metric, will automatically be parallel-transported.
Consider the full contraction:
\ba 
F_{abc}H^{abc}&=&\frac{1}{9}(u_a h_{bc}+u_c h_{ab}+u_b h_{ca})\nonumber\\
&&\quad \times(u^a k^{bc}+u^c k^{ab}+u^b k^{ca})\,\nonumber\\
&=&\frac{1}{9}\Bigl(3 u^2 (h \cdot k)-2u^au^b (h\cdot k)_{ba}\nonumber\\
&&\quad -2u^au^c (h\cdot k)_{ac}-2 u^cu^b(k\cdot h)_{cb}\Bigr)\nonumber\\
&=&
\frac{1}{9}\Bigl(3 g_{ab} (h \cdot k)-6(h\cdot k)_{ab}\Bigr)u^au^b\,, 
\ea
where we have abbreviated 
\be 
(h\cdot k)_{ab}\equiv h_{ac}k_b{}^c\,,\quad h\cdot k\equiv h_{ab}k^{ab}\,.
\ee 
Since the above equation is of the form \eqref{wab}, we conclude that (normalizing the second term to unity), the following object:
\be 
Q_{ab}=(h\cdot k)_{(ab)}-\frac{1}{2}g_{ab}(h \cdot k)\,,
\ee 
is a Killing tensor. Note also that the second term vanishes for null geodesics, $u^2=0$. This means that 
\be 
Q_{ab}=(h\cdot k)_{(ab)}
\ee 
is a conformal Killing tensor.

{Consider next the following contraction: 
\ba 
9F_{acd}H_b{}^{cd}&=&u_au_b (h \cdot k)+2u^2 (h \cdot k)_{ab}-2u_au^d(k\cdot h)_{bd}\nonumber\\
&&-2u_b u^c (h \cdot k)_{ac}+u^d u^c(k_{db}h_{ac}+h_{da}k_{bc})\nonumber\\
&=&u^cu^d\Bigl(
g_{ca}g_{db}(h \cdot k)
+2(h \cdot k)_{ab}g_{cd}\nonumber\\
&&\quad -2g_{ac}(k\cdot h)_{bd}-2g_{db}(h\cdot k)_{ac}
-2h_{ca}k_{db}\Bigr)\nonumber\\
&\equiv&Q_{cadb}u^cu^d\,.
\ea 
Since this is again precisely of the form \eqref{wab}, we may read off the following representative GKT (normalizing the last term to unity):
\ba 
Q_{abcd}&=&h_{ab}k_{cd}+(k \cdot h)_{ab}g_{cd}+g_{ab}(h \cdot k)_{cd}\nonumber\\
&&-\frac{1}{2}g_{ab}g_{cd}(h \cdot k)-g_{ac}(h\cdot k)_{bd}\,.
\ea
}
Again, when the geodesic is null, $u^2=0$, the last term vanishes, and we have a CGKT
{\ba 
\tilde Q_{abcd}&=&h_{ab}k_{cd}+(k \cdot h)_{ab}g_{cd}+g_{ab}(h \cdot k)_{cd}\nonumber\\
&&-\frac{1}{2}g_{ab}g_{cd}(h \cdot k)\,,
\ea }
obeying \eqref{CGKT}. We can easily identify the additional terms when compared to \eqref{kkill}.

A generalization to closed conformal KY $p$-forms $h$ and $k$ is now straightforward.
We define the corresponding parallel-transported  $(p+1)$-forms  
$F$ and $H$ and consider the following contraction: 
\be \label{conj1}
F_{aa_1\dots a_{p}}H_b{}^{a_1\dots a_p}\equiv Q_{cadb}u^cu^d\,.
\ee 
Performing the calculation, see Appendix~\ref{AppB}, we recover the following formula for the GKT: 
\ba\label{QKKThh}
    Q_{abcd} &=& h_{ab c_2 \dots c_p} k_{cd}{}^{c_2 \dots c_p}+\tfrac{1}{p-1} g_{ab}(h\cdot k)_{cd} \nonumber\\
    &&+\tfrac{1}{p-1} g_{cd}(h\cdot k)_{ba} -\tfrac{1}{p-1}g_{ac}(h\cdot k)_{bd}\nonumber\\
    &&-\tfrac{1}{p(p-1)}g_{ab}g_{cd} (k\cdot h)\,,
\ea
where we have defined 
\be 
(k \cdot h)_{ab}=k_{aa_2\dots a_p}h_b{}^{a_2\dots a_p}\,,\quad (h \cdot k)=h_{a_1\dots a_p}k^{a_1\dots a_p}\,.
\ee 
This then yields the following CGKT:
\ba\label{CGKTff22} 
\tilde Q_{abcd}&=& h_{ab c_2 \dots c_p} k_{cd}{}^{c_2 \dots c_p}+\tfrac{1}{p-1} g_{ab}(h\cdot k)_{cd} \nonumber\\
    &&+\tfrac{1}{p-1} g_{cd}(h\cdot k)_{ba}-\tfrac{1}{p(p-1)}g_{ab}g_{cd} (k\cdot h)\,.\qquad
\ea

Having established \eqref{CGKTff22}, it is now obvious that the general Kerr-NUT-AdS spacetimes presented in the previous section 
provide an example of spacetimes 
admitting conformal generalizations of GKTs defined in \eqref{CGKT}.
Namely, considering the closed conformal KY $(2p)$-forms $h_p$, given by \eqref{hp}, we have the associated CGKTs 
\eqref{CGKTff22}:
\ba \label{CGKTff22b}
\tilde Q_{abcd}^{(p)}&=& (h_p)_{ab a_3 \dots a_{2p}} (h_p)_{cd}{}^{a_3 \dots a_{2p}}
+\tfrac{1}{2p-1}(h_p^2)_{ab}g_{cd}\nonumber\\
&&
+\tfrac{1}{2p-1} g_{ab}(h_p^2)_{cd}-\tfrac{1}{2p(2p-1)}g_{ab}g_{cd}(h_p^2)\,.
\ea 
The purpose of these tensors in these spacetimes remains to be investigated.

Of course, one can also straightforwardly  extend these results to higher-rank (C)GKTs, obtained as products of closed conformal KY tensors `with fewer contractions'.

\subsection{Conformal property}
It is well known that 
under conformal transformations 
\be \label{Omega2g}
g_{ab}\to \hat g_{ab}=\frac{1}{\Omega^2}g_{ab}\,,\quad g^{ab}\to \hat g^{ab}=\Omega^2 g^{ab}\,,
\ee 
the conformal Killing tensors transform as 
\be\label{conformalProp} Q^{ab}\to \hat Q^{ab}=Q^{ab}\,.
\ee

{In its turn, the conformal property \eqref{conformalProp} together with the Killing tensor equation \eqref{KT} can be used to `motivate' the proper conformal generalization 
of Killing tensors. To show this we proceed as follows. 
Consider a frame where we have a full Killing tensor, $Q^{ab}$ obeying 
\be
\nabla^{(a}Q^{bc)}=0\,.
\ee 
Performing the conformal transformation,  
we find that (not assuming any symmetry for $Q^{ab}$):
\begin{align}\label{2formconfnonsym}
    \hat\nabla^{(a}\hat Q^{bc)}=&\nabla^{(a}Q^{bc)}\nonumber\\
    &-\hat{g}^{(ab}Q^{c)d}\Omega\nabla_d\Omega-\hat{g}^{(ab}Q^{|d|c)}\Omega\nabla_d\Omega\,,
\end{align}
where we have used \eqref{Omega2g}, \eqref{conformalProp}, and the formula for covariant derivative of the conformally scaled metric, namely
\be\label{Gamma}
\hat{\Gamma}^{c}_{ab}-\Gamma^{c}_{ab}
=
-\delta^{c}_{a}\Upsilon_{b}
-\delta^{c}_{b}\Upsilon_{a}
+g_{ab}\Upsilon^{c},
\quad
\Upsilon_{a}\equiv\frac{\nabla_{a}\Omega}{\Omega}\,.
\ee
Since $Q^{ab}$ is a Killing tensor w.r.t. the original $\nabla_a$, we thus find that in the new frame we have  
\be 
\hat \nabla^{(a}\hat Q^{bc)}=\hat g^{(ab}\xi^{c)}\,,
\ee 
where $\xi^c$ is some vector, in our case given by 
$\xi^c=Q^{cd}\Omega \nabla_d \Omega+Q^{dc}\Omega \nabla_d \Omega$. We have just `derived' the form of the conformal Killing tensor equation. 
}

{Let us now proceed similarly with the GKTs. Since we know that $Q^{abcd}$ yields a (not necessarily symmetric) Killing tensor by $Q^{ab}=Q^{acbd}g_{cd}$, it means that if CGKTs are to have a conformal property, they must transform as 
\be \label{QabcdConf}
Q^{abcd}\to \hat Q^{abcd}=\Omega^2 Q^{abcd}\,,
\ee 
under conformal transformations \eqref{Omega2g}.
So starting again with a GKT $Q^{abcd}$ in one frame 
\be\label{Q1frame} 
\nabla^{(a}Q^{b|c|d)e}=0\,,
\ee 
and transforming to a conformally related frame \eqref{Omega2g}, 
using \eqref{QabcdConf} and \eqref{Gamma}, we find 
\ba 
\hat \nabla^{(a}\hat Q^{b|c|d)e} &=& \Omega^4 \nabla^{(a}Q^{b|c|d)e}\nonumber\\
&&+\Upsilon_{f}\Bigl(-\hat{g}^{(ab}\hat{Q}^{d)cfe}
-\hat{g}^{(ab}\hat{Q}^{f c|d)e}
\nonumber\\
&&\  -\hat{g}^{c(a}\hat{Q}^{b|f|d)e}
-\hat{g}^{e(a}\hat{Q}^{b|c|d)f}
\nonumber\\
&&\  +\hat g^{cf}\hat{Q}^{(abd)e} +\hat g^{ef}\hat{Q}^{(a|c|bd)} \Bigr)\,.\quad 
\ea 
Obviously, the first term vanishes due to \eqref{Q1frame}. The second line then has the form required by \eqref{CGKT}. However, the third and fourth lines do not necessarily disappear unless some additional restrictions are imposed on $\hat Q_{abcd}$. This seems to indicate that general CGKTs \eqref{CGKT} do not seem to have the conformal property \eqref{QabcdConf}, although their conformal Killing tensor contractions do. Whether a subset of CGKTs can be defined which would preserve conformal property remains to be seen.     
}

\subsection{Conformal Killing--Yano squares}

Let us finally return to the question of whether conformal KY tensors square to CGKTs. We consider the case of null geodesics 
\be 
l^2=0\,,\quad \nabla_l l^a=0\,,
\ee  
and for simplicity, focus on the case of a single CKY 2-form $k$.

{Then we know that the following 2-form: 
\be 
H=l\wedge (l\cdot k)\,,
\ee 
or 
\be 
H_{ab}=l_{[a}k_{|c|b]}l^c\,,
\ee 
is parallel-transported along null geodesics,
\be 
\nabla_l H=0\,.
\ee 
see \eqref{Hformula}--\eqref{Hcomps} in Appendix~\ref{AppA}.
}

Since $H_{ab}l^b=0$, the following product identically vanishes:
\be 
H_{ab}H^{ab}=0\,.
\ee 
Considering next
\ba 
H_{ac}H_b{}^c&=&\frac{1}{4}\bigl(l_a k_{dc}l^d-l_c k_{da}l^d\bigr)\bigl(l_b k^{ec}l_e-l^ck^e{}_bl_e\bigr)\nonumber\\
&=&-\frac{1}{4}l_al_b Q\,,\quad Q\equiv (k^2)_{de}l^dl^e\,.
\ea 
Since $l_a$ and $l_b$ are parallel-transported, so must be $Q$. This shows that 
\be 
Q_{ab}=(k^2)_{ab}
\ee 
is a conformal Killing tensor.

Consider finally the non-contracted product
\ba \label{HHbad}
4H_{ab}H_{cd}&=&
\bigl(l_a k_{eb}l^e-l_b k_{ea}l^e\bigr)\bigl(l_c k_{fd}l^f-l_dk_{fc}l^f\bigr)\nonumber\\
&=&l_a l_c W_{bd}-l_al_d W_{bc}-l_bl_c W_{ad}+l_bl_dW_{ac}\,\nonumber\\
&=&2l_al_{[c}W_{|b|d]}-2l_bl_{[c}W_{|a|d]}\,,
\ea 
where we have abbreviated
\be \label{Wbad}
W_{bd}\equiv k_{ab}k_{cd}l^al^c=W_{db}\,.
\ee 
Unfortunately, this does not seem to lead to a rank-4 CGKT. Rather, one can rewrite \eqref{HHbad} as 
\ba 
4 H_{ab}H_{cd}&=&\bigl(g_{az}g_{by}W_{bd}-g_{az}g_{dy}W_{bc}\nonumber\\
&&-g_{bz}g_{cy}W_{ad}+g_{bz}g_{dy}W_{ac}\bigr)l^zl^y\,.
\ea 
Unpacking $W$ using \eqref{Wbad} then leads to rank-8 CGKT. Unfortunately, we currently do not know if a rank-4 CGKT can be constructed from $k$.

\section{Discussion}
\label{Sec6}

In this paper, we have uncovered a very rich structure of higher-rank (generalized) hidden symmetries that may play a fundamental role 
for parallel transport and the separability of test field equations  
in curved spacetimes.

To start with, we  have generalized the construction of Killing tensors of Collinson and Howarth \cite{collinson2000generalized} by relaxing the condition on the complete symmetry of parallel-transported indices, which, when contracted with the particle's momenta, yield parallel-transported objects. This gives rise to equivalence classes of generalized (hidden) symmetries encoded in GKT representatives 
that agree upon the complete symmetrization of chosen indices and  yield the same 
parallel-transported objects. Different representatives of a given class may possess various extra symmetries. Especially useful seems to be a generalization of rank-(2-2) GKTs to `Riemann-type' GKTs with additional Riemann-like symmetries. Such objects naturally arise as partially contracted squares of KY tensors; they play an interesting role in $p$-form generalizations of Proca-type theories \cite{Kubiznak:2026snm} and happen to exist in (higher-dimensional) rotating black hole spacetimes.

We have also proposed conformal-like generalizations of such tensors, which we call CGKTs.
We have shown that such objects give rise to parallel-transported tensors along null geodesics and naturally appear in Kerr-NUT-AdS spacetimes, where they arise as `upgraded' partially contracted  squares of  closed conformal KY tensors; see Eqs.~\eqref{CGKTff22} and \eqref{CGKTff22b}. Whether such a construction can be generalized to a weaker structure of conformal KY tensors has only been partially addressed and will be studied elsewhere. We have also shown that, contrary to standard conformal Killing tensors, the new objects do not seem to (at least in general) transform nicely under conformal transformations.

Our construction raises many interesting questions. First, it is well known that 
(standard) Killing tensors form an algebra w.r.t. the so called Schouten--Nijenhuis brackets. Namely, having a rank $p$ and rank $q$ Killing tensors $K_p$ and $K_q$, their SN bracket, defined by 
\ba 
[K_p,K_q]_{\mbox{\tiny SN}}^{a_1\dots a_{p+q-1}}&\equiv& p 
K_p^{c(a_1\dots a_{p-1}} \nabla_c K_q^{a_p\dots a_{p+q-1})}\nonumber\\
&&-qK_q^{c(a_1\dots a_{q-1}}\nabla_c K_p^{a_q\dots a_{p+q-1})}\,,\qquad
\ea 
is again a Killing tensor.\footnote{This property has recently been used to construct physically well motivated spacetimes with irreducible higher-rank Killing tensors \cite{Gray:2021toe}.}       
Is it possible to generalize the SN bracket in such a way that the GKTs would form a similar type of algebra?

Second,
we have defined the above GKTs based on their property of giving parallel-transported tensors along geodesics. It remains to be studied whether such objects, similar to what happens with standard Killing and KY tensors, e.g., \cite{Carter:1977pq, Carter:1979fe, Andersson:2014lca, Frolov:2017kze}, also play any fundamental role in the {\em separability} of (possibly higher spin) test field equations and in the construction of the corresponding {\em symmetry operators}. In particular,   
although the Killing objects in a given class give rise to the same parallel-transported objects along geodesics, they may possess different types of additional symmetries. It may be the case that for the construction of symmetry operators for a given test field equation, some representatives are `more useful' than others.

Third,
in our study, we have mostly focused on rank-(2-2) generalizations of (conformal) Killing tensors. Such objects naturally exist in higher-dimensional black hole spacetimes. However, other rank objects may also be useful. For example, one may consider a rank-(1-2) GKT $Q_{abc}$ obeying 
\be 
\nabla_{(a} Q_{|b|cd)}=0\,.
\ee 
Whether such `vector-valued' rank-2 Killing tensors exist in some (physically motivated) geometries remains to be seen.\footnote{A reducible example of such an object can trivially be obtained by taking a product of a KY 2-form with a Killing vector.}

Fourth, it is now well established that there are some useful extensions of KY tensors. For example, the `torsion generalization' of KY tensors proves to be useful in string theory and SUGRA motivated spacetimes, e.g. \cite{Kubiznak:2009qi, Houri:2010fr}. It is also known that such a `weaker' KY structure gives rise to standard Killing tensors. A possible direction of study is to see whether the same remains true for the GKTs.

It seems that a very rich structure of possible hidden symmetries in curved spacetimes still remains to be uncovered.


\appendix

\subsection*{Acknowledgements}

D.K. and C.A. acknowledge support from the Charles University Research Center Grant No. UNCE24/SCI/016.

\section{Notes on conformal Killing--Yano tensors}
\label{AppA}

Let us gather here  a basic overview of the conformal KY tensors used in the main text. We mostly follow \cite{Frolov:2017kze}.

\subsection{Definitions and basic properties}

A {\em conformal Killing--Yano (CKY)} $p$-form $k$ is defined by the following equation: 
\be\label{CKYdef} 
\nabla_X k=X\wedge \xi+  X\cdot \kappa\,,
\ee 
where $X$ is an arbitrary vector field. It follows that 
\be 
\xi=\frac{1}{D-p+1}\nabla \cdot k\,,\quad \kappa=\frac{1}{p+1}dk\,,
\ee 
or in components:
\be 
\nabla_a k_{a_1\dots a_p}=
\nabla_{[a} k_{a_1\dots a_p]}+\frac{p}{D-p+1}g_{a[a_1}\nabla^b{}k_{|b|a_2\dots a_p]}\,.
\ee 
Such objects behave nicely under a conformal transformation. Namely, when
\be 
g \ \to\  \hat g\equiv\Omega^{-2} g\,,
\ee 
then having a CKY $p$-form $k$ for $g$,
\be 
k \ \to \ \hat k\equiv \frac{1}{\Omega^{p+1}}k
\ee 
preserves the CKY property. The definition is also invariant under a Hodge duality. Namely, whenever $k$ is a CKY $p$-form, then 
$(*k)$
is a CKY ($D-p)$-form.

{\em Killing--Yano (KY) tensors} are a subset of CKY for which the divergence part vanishes. That is, a KY $p$-form $f$ obeys
\be\label{KYdef} 
\nabla_X f =X \cdot \kappa\,,
\ee 
or 
\be 
\nabla_a f_{a_1\dots a_p}=
\nabla_{[a} f_{a_1\dots a_p]}\quad \Leftrightarrow \quad \nabla_{(a} f_{a_1)\dots a_p}=0\,.
\ee 

{\em Closed conformal Killing--Yano (CCKY)} tensors are a subset of CKY tensors with a vanishing exterior part. That is, a $p$-form $h$ is a CCKY tensor when it obeys
\be\label{CCKYdef} 
\nabla_X h=X\wedge \xi\,,
\ee 
or 
\be \label{CCKYdef2}
\nabla_a h_{a_1\dots a_p}=
\frac{p}{D-p+1}g_{a[a_1}\nabla^b{}h_{|b|a_2\dots a_p]}\,.
\ee 
The following important property was shown in \cite{Krtous:2006qy}. When $h_1$ and $h_2$ are two CCKY tensors, so is their exterior product
\be 
h_1\wedge h_2\,.
\ee 

Finally, we note that under Hodge duality, KY tensors map to CCKY tensors and vice versa.

\subsection{Parallel transport}

Let us first consider the case of timelike geodesics:
\be 
\nabla_u u^a=0\,.
\ee 
Then, having a KY $p$-form $f$, we may define a $(p-1)$-form 
\be\label{w_KY} 
w=u \cdot f\quad \Leftrightarrow \quad w_{a_2\dots a_p}=u^a f_{aa_2\dots a_p}\,.
\ee 
Such a form is then automatically parallel-transported along the above timelike geodesics. Indeed, we have 
\be 
\nabla_u w= u \cdot (\nabla_u f)=u \cdot (u \cdot \kappa)=0\,,
\ee 
where in the second equality, we have used the KY defining property \eqref{KYdef}. 

Similarly, having a CCKY $p$-form $h$, we may define a parallel-transported $(p+1)$-form 
\be\label{FCCKY} 
F=u \wedge h\quad \Leftrightarrow\quad  F_{aa_1\dots a_p}=u_{[a}h_{a_1\dots a_p]}\,.
\ee 
Indeed, we have 
\be 
\nabla_u F= u \wedge (\nabla_u h)=u \wedge (u \wedge \xi)=0\,,
\ee 
where we have used \eqref{CCKYdef}. So both KY and CCKY tensors give rise to parallel-transported objects along timelike geodesics. We believe that the second observation is new.
Of course, both of the above results also remain true for the null geodesics
\be 
\nabla_l l^a=0\,,\quad l^2=0\,.
\ee

When only a CKY tensor is present, the situation is significantly more complicated. We focus on the null case. Let $k$ be a CKY $p$-form. Defining
\be 
\hat H\equiv l\wedge k +l \cdot \beta\,,
\ee 
for some $(p+2)$-form $\beta$, we then have 
\ba 
\nabla_l \hat H&=&l \wedge (\nabla_l k)+l \cdot (\nabla_l \beta)\nonumber\\
&=&
l \wedge (l \wedge \xi+l \cdot \kappa)+l \cdot (\nabla_l \beta)\nonumber\\
&=&l \wedge (l \cdot \kappa)+l \cdot (\nabla_l \beta)\nonumber\\
&=&-l \cdot (l \wedge \kappa)+l \cdot (\nabla_l \beta)\nonumber\\
&=&l \cdot (\nabla_l \beta-l \wedge \kappa)\,, \label{divH}
\ea 
where we have used the defining CKY property \eqref{CKYdef}, together with the fact that $l$ is null. 
Thus, $\hat H$ is parallel-transported, provided we choose $\beta$ {which obeys
\be 
\nabla_l \beta=l\wedge \kappa+l\cdot \tilde \beta
\ee 
for some $(p+3)$-form $\tilde \beta$. 
For a given $l$, this is a differential equation for $\beta$ that, in principle, can be solved. Note, however, that if this is required for any $l$, it leads to the conclusion that $\beta$ is a {$(p+2)$} CKY tensor, with a constraint that 
\be 
\kappa=\frac{1}{p+1}dk=\frac{1}{D-(p+2)+1}\nabla \cdot \beta\,.
\ee 
This means that 
$\kappa$ is both closed, $d\kappa=0$, and co-closed, $\nabla \cdot\kappa =0$, thus it is harmonic, $\triangle\kappa = 0$. This seems to be a rather restrictive requirement on the properties of the manifold. \footnote{On a well-behaved (closed Riemannian) manifold, this would also mean that $\kappa = 0$, forcing $\nabla\cdot \beta=0$ and $k$ to be a CCKY.}

Similarly, we may consider a different form
\be 
\tilde H= l \cdot k +l \wedge \gamma\,,
\ee 
for some $(p-2)$-form $\gamma$. By the same argument as in \eqref{divH}, we obtain 
\ba 
\nabla_l \tilde H&=&
l \cdot (l \wedge \xi+l \cdot \kappa)+ l \wedge \nabla_l\gamma\nonumber\\
&=&l \wedge (\nabla_l\gamma- l \cdot \xi)\,.
\ea 
Thus, it is parallel-transported when $\gamma$ {obeys 
\be 
\nabla_l \gamma=l \cdot \xi+l\wedge \tilde \gamma\,.
\ee 
If this is demanded for any $l$, it means that $\gamma$ has to be a CKY $(p-2)$-form,} with an additional constraint that 
\be 
\xi=\frac{1}{D-p+1} \nabla \cdot k=\frac{1}{p-1} d\gamma\,.
\ee 

To avoid the necessity of finding $\beta$ or $\gamma$, we may upgrade the above forms $\hat H$ and $\tilde H$ by considering 
\ba
l \cdot \hat H&=&l \cdot (l \wedge k+l \cdot \beta)=l \cdot (l \wedge k)\nonumber\\
&=&-l \wedge (l \cdot k)=-l \wedge (l \cdot k+l \wedge \gamma)\nonumber\\
&=&-l \wedge \tilde H\,.
\ea
In other words, for a CKY $p$-form $k$, we may assign a $p$-form 
\be 
{H\equiv l\wedge (l \cdot k)=-l \cdot (l \wedge k)\,.} \label{Hformula}
\ee 
or in components 
\be 
H_{a_1\dots a_p}=l_{[a_1}k_{ba_2\dots a_p]}l^b\,,\label{Hcomps}
\ee 
which is automatically parallel-transported along null geodesics $l$. Note, however, that the important difference between the forms $w$ and $F$ on one hand, and the form $H$ on the other, is that the latter is now quadratic instead of linear in the geodesic velocity. 
{In addition, the $p$-form $H$ is also degenerate, as we obviously have 
\be 
l \cdot H=0=l\wedge H\,.
\ee
}

\section{Construction of conformal GKT from CCKY $p$-form contractions}\label{AppB}

In Appendix ~\ref{AppA}, we have seen that CCKY $p$-forms give rise to parallel-transported $(p+1)$-forms \eqref{FCCKY}. Here, we shall use this property to find conformal rank-(2-2) GKTs as `squares' of CCKY $p$-forms.

Let $h$ and $k$ be two CCKY $p$-forms. Then, according to \eqref{FCCKY}, the following ($p+1$)-forms are parallel-transported along any geodesic $u^a$:
\begin{equation}
    F_{ac_1 \dots c_p} \equiv u_{[a}h_{c_1\dots c_p]}\,,\quad K_{ac_1 \dots c_p} \equiv u_{[a}k_{c_1\dots c_p]}\,.
\end{equation}
Note that we can rewrite
\begin{align}
    u_{[a}h_{c_1\dots c_p]} =& \tfrac{1}{p+1}(u_a h_{c_1\dots c_p}-\sum_{i=1}^{p}u_{c_i}h_{c_1\dots c_{i-1}ac_{i+1}\dots c_p})\nonumber\\
    =&\tfrac{1}{p+1}(u_a h_{c_1\dots c_p}+\sum_{i=1}^{p}(-1)^iu_{c_i}h_{a \hat C_i})\,,
\end{align}
where $\hat C_i$ denotes the range $\{c_1,\dots,c_p\}$ without $c_i$.
Consider now the following parallel-transported object: \eqref{conj1}: 
\begin{align}
    I_{ab} \equiv& F_{a c_1 \dots c_p}K_b{}^{c_1\dots c_p}\nonumber\\
    =& \tfrac{1}{(p+1)^2}(u_a h_{c_1\dots c_p}+\sum_{i=1}^{p}(-1)^iu_{c_i}h_{a \hat C_i})\nonumber\\
    &\quad\cdot(u_b k^{c_1\dots c_p}+\sum_{j=1}^{p}(-1)^ju^{c_j}k_{b}{}^{\hat C_j})\nonumber\\
    \equiv&\tfrac{1}{(p+1)^2}(u_a u_b (k\cdot h) + A_{ab}+B_{ab}+C_{ab})\,,
\end{align}
where we denoted the structurally different terms by auxiliary tensors. Namely, 
\ba
A_{ab} &\equiv& u_a h_{c_1 \dots c_p}\sum_{j=1}^{p}(-1)^ju^{c_j}k_{b}{}^{\hat C_j}\nonumber\\
&=&-p u_a (u\cdot h)_{\hat C_1}k_b{}^{\hat C_1}\,,
\ea
where we commuted the contraction over $c_j$ to the first position, yielding $(-1)^{j-1}$, and renamed indices.
In the same fashion, we have
\ba
    B_{ab}&\equiv& \Bigl(\sum_{i=1}^{p}(-1)^iu_{c_i}h_{a \hat C_i}\Bigr) u_b k^{c_1 \dots c_p} \nonumber\\
    &=&-p h_{a\hat C_1} u_b (u \cdot  k)^{\hat C_1}\,.
\ea
The last term reads
\begin{align}
    C_{ab} \equiv \sum_i \sum_j (-1)^{i+j} u_{c_i}u^{c_j} h_{a \hat C_i} k_{b}{}^{\hat C_j}\,.
\end{align}
We split the sum into the `diagonal' sum $i=j$ of $p$ equivalent terms, and the off-diagonal $i \neq j$ of $(p^2-p)$ terms. For the diagonal terms, we get
\be
C_{ab}^{\mathrm{diag}} \equiv p u^2 h_{a\hat C_1}k_{b}{}^{\hat C_1}\,, 
\ee
since the factor $(-1)^{i+j} = 1$. For the second term, we commute the contraction to the first index for both $u$'s. Regardless of whether $i<j$ or $i>j$, the resulting factor is $(-1)$, and we are left with
\begin{equation}
    C_{ab}^{\mathrm{non-diag}} = -p(p-1) (u\cdot h)_{a c_3\dots c_p} (i_u k)_{b}{}^{c_3\dots c_p}\,.
\end{equation}
Putting everything together, we have  
\ba
    I_{ab} &=& \tfrac{1}{(p+1)^2}\left[ u_a u_b (k\cdot h) -p u_a (u\cdot h)_{\hat C_1}k_b{}^{\hat C_1}\right.\nonumber\\
    &&-p h_{a\hat C_1} u_b (u\cdot k)^{\hat C_1}+p u^2 h_{a\hat C_1}k_{b}{}^{\hat C_1}\nonumber\\
    &&\left.-p(p-1) (u\cdot h)_{a  c_2\dots c_p} (u\cdot k)_{b}{}^{ c_2\dots c_p}\right]\nonumber\\
    &=& \tfrac{1}{(p+1)^2}u^cu^d\Bigl(g_{ca}g_{db} (k\cdot h) - p g_{ca} h_{d\hat C_1}k_b{}^{\hat C_1}\nonumber\\
    &&-p g_{db}h_{a\hat C_1} k_{c}{}^{\hat C_1} + p g_{cd}h_{a\hat C_1}k_b{}^{\hat C_1}\nonumber\\
    &&-p(p-1) h_{ca c_2 \dots c_p} k_{db}{}^{c_2 \dots c_p}\Bigr)\nonumber\\
    &\equiv&-\tfrac{p(p-1)}{(p+1)^2}u^cu^d Q_{cadb}\,,
\ea
giving the following GKT  representative:
\ba\label{QAppB}
    Q_{abcd} &=& h_{ab c_2 \dots c_p} k_{cd}{}^{c_2 \dots c_p}+\tfrac{1}{p-1} g_{ab}(h\cdot k)_{cd} \nonumber\\
    &&+\tfrac{1}{p-1} g_{cd}(h\cdot k)_{ba} -\tfrac{1}{p-1}g_{ac}(h\cdot k)_{bd}\nonumber\\
    &&-\tfrac{1}{p(p-1)}g_{ab}g_{cd} (k\cdot h)\,,
\ea
which is the formula \eqref{QKKThh} used in the main text.


%

\end{document}